\documentclass[a4paper,12pt]{article}
\usepackage{latexsym,amssymb}
\usepackage{graphicx}
\usepackage{cite}
\usepackage[arrow,matrix,curve]{xy}
\usepackage{amsmath}

\usepackage{amsmath}
\usepackage{amssymb}
\usepackage{amscd}
\usepackage{amsfonts}
\usepackage{amsthm}
\usepackage[matrix,arrow,curve]{xy}

\global\arraycolsep=2pt 
\input{epsf} 

\def\s[#1,#2]{[#1\stackrel{{\displaystyle\star}}{,}#2]}

 1

\usepackage{amsthm}

\newcommand{\eq}{\begin{equation}}
\newcommand{\eqa}{\begin{eqnarray}}
\newcommand{\en}{\end{equation}}
\newcommand{\ena}{\end{eqnarray}}
\newcommand{\enn}{\nonumber \end{equation}}

\def\sk{\vskip .4cm}
\def\noi{\noindent}

\def\de{\delta}
\def\epsi{{\varepsilon}}
\def\st {\star}

\def\of{{\overline{{\rm{f}}\,}}}

\def\D/h{\widehat{\fmslash D}}

\def\al{\alpha}
\def\la{\lambda}
\def\be{\beta}
\def\ga{\gamma}

\def\de{\delta}

\def\5bar{{\overline 5}}

\def\R{{R}}

\def\FF{\mathcal F}

\def\s'O{\stackrel{_{{\displaystyle\st \footnotesize '}}}{_{^{^{\displaystyle\otimes}}}}}

\def\La{\Lambda }

\def\D{\Delta}
\def\1s{{1_\st }}
\def\3s{{3_\st }}
\def\2s{{2_\st }}
\def\ef1{{1_\FF}}
\def\ef2{{3_\FF}}
\def\ef3{{2_\FF}}

\def\le{\langle}
\def\re{\rangle}
\def\nn{\nonumber}

\def\cc{\mathbb{C}}  
\def\rr{\mathbb{R}}  
\def\r4{\mathbb{R}^4}  
\def\UP{U(iso(3,1))}

\def\T#1#2{ T^{#1}_{~~#2} }
\def\Ti#1#2{ (T^{-1})^{#1}_{~~#2} }

\def\TP{ T^{\prime} }

\def\qm{q^{-1}}

\def\c#1#2{ C_{#1}^{~~~#2} }

\def\U{U_q(sl(2))}
\def\mdot{\mu}
\def\gp{g'}
\def\kappa{S}
\def\Fun{$Fun(G)$ }
\def\R#1#2{ R^{#1}_{~~~#2} }
\def\qonelim{\stackrel{q \rightarrow 1}{\longrightarrow}}

\def\Lpm#1#2{{L^{\pm}}^{#1}_{~~#2}}
\def\LLpm{L^{\pm}}
\def\LLp{L^{+}}
\def\LLm{L^{-}}
\def\Lp#1#2{{L^{+}}^{#1}_{~~#2}}
\def\Lm#1#2{{L^{-}}^{#1}_{~~#2}}

\def\R#1#2{ R^{#1}_{~~~#2} }

\def\R#1#2{ R^{#1}_{~~~#2} }

\numberwithin{equation}{section}

\begin{document}

\begin{titlepage}
\rightline{DISTA-UPO/07}
\sk\sk\sk\sk
\begin{center}
{\bf\LARGE{Lectures on Hopf Algebras,\\[.2em] Quantum Groups and Twists$\,^{^{_{_\st}}}$}}\\[.5em] 
\vskip 2.5em

{{\bf Paolo Aschieri}}

\vskip 1.5em
Centro Studi e Ricerche Enrico Fermi, Compendio Viminale, I-00184, Roma, Italy

Dipartimento di Scienze e Tecnologie Avanzate\\
Universit\' a del Piemonte Orientale, and INFN - Sezione di Torino\\
Via Bellini 25/G 15100 Alessandria, Italy\\[1em]
\end{center}

\sk
\sk
\sk
\centerline{\bf Abstract}
\sk
\normalsize{
Lead by examples we introduce the notions of Hopf algebra
and quantum group. We study their geometry and in particular their
Lie algebra (of left invariant vectorfields). The examples of the quantum $sl(2)$ Lie algebra and of 
the quantum (twisted) 
Poincar\'e Lie algebra 
$iso_\theta(3,1)$ are presented.}

\sk\sk
\sk\sk

\sk\sk
\noi

\sk\sk
\noi{$^\st\,$Lectures given at the second Modave Summer School in Mathematical Physics, August 6-12, 2006.}

\sk
\noi {\footnotesize{E-mail: aschieri@mfn.unipmn.it}}\\

\sk\sk


\end{titlepage}\vskip.2cm

\newpage

\section{Introduction}
Hopf algebras where initially considered more than Half a century ago. New 
important examples, named quantum groups, where studied in the 80's 
\cite{Drinfeld, Jimbo, FRT}, they arose in the study of the quantum inverse scattering method in integrable systems.
Quantum groups can be seen as symmetry groups of noncommutative spaces, this is one reason they have been investigated in physics and mathematical physics
(noncommutative spaces arise as quantization of commutative ones).
The emergence of gauge theories on noncommutative spaces
in open string theory in the presence 
of a NS 2-form background \cite{SW}
has further motivated the study of noncommutative 
spaces, and of their symmetry properties. 

We here introduce the basic concepts of quantum group and of its  
Lie algebra of infinitesimal transformations. We pedagogically stress 
the connection with the classical (commutative) case and we treat two 
main examples, the quantum $sl(2)$ Lie algebra and the quantum Poincar\'e 
Lie algebra. 
\sk
Section 2 shows how commutative Hopf algebras emerge from groups. 
The quantum group $SL_q(2)$ is then presented and its corresponding 
universal enveloping algebra $U_q(sl(2))$ discussed. The relation between
$SL_q(2)$ and $U_q(sl(2))$ is studied in Section 5. The quantum $sl(2)$
Lie algebra, i.e. the algebra of infinitesimal transformations, is then studied 
in Section 6. Similarly the geometry of Hopf algebras obtained from 
(abelian) twists is studied via the example of the Poincar\'e  Lie algebra.

In the appendix for reference we review some basic algebra notions and define 
Hopf algebras diagramatically.
\sk
One aim of these lectures is to concisely introduce and relate all three 
aspects of quantum groups: 
\begin{itemize}
\item {deformed algebra of functions \cite{FRT},} 
\item{deformed universal enveloping algebra \cite{Drinfeld, Jimbo},}
\item{quantum Lie algebra \cite{Woronowicz}.}
\end{itemize} 
Quantum Lie algebras encode the 
construction of the (bicovariant) differential calculus and 
geometry, most relavant for physical applications.
A helpful review for the  first and second aspects is \cite{Majid},  
for quantum Lie algebras we refer to \cite{Zumino} and 
\cite{AC}. The (abelian) twist case, that 
is an interesting subclass, can be found in \cite{GR2} and in
\cite{Aschieri2006}.

\section{Hopf algebras from groups}
Let us begin with two examples motivating the notion of Hopf algebra.
Let $G$ be a finite group,
and $A=Fun(G)$ be the set of functions from $G$ to
complex numbers $\cc$. $A=$\Fun is an algebra over $\cc$ with the usual  
sum and product $(f+h)(g)=f(g)+h(g),~(f \cdot h)=f(g) h(g),
~(\lambda f)(g)=\lambda f(g)$, for $f,h \in Fun(G),~g \in G,~\lambda \in 
\cc$. The unit of this algebra is $I$, defined by $I(g)=1,~\forall g \in 
G$. 
Using the group structure of $G$ (multiplication map, existence of unit element and inverse element), we can introduce on \Fun three other 
linear maps, the coproduct $\D$, the counit $\epsi$, and 
the coinverse (or antipode) $\kappa$:
\begin{eqnarray}
\D (f)(g,\gp) &\equiv& f(g\gp),~~~\D:Fun(G) \rightarrow Fun(G)\otimes 
  Fun(G) \label{cop} \\
\epsi(f) &\equiv& f(1_G),~~~~~~\epsi:Fun(G) \rightarrow \cc \label{cou}\\
(\kappa f)(g) &\equiv& f(g^{-1}),~~~\kappa:Fun(G) \rightarrow Fun(G)
  \label{coi}
\end{eqnarray}
\noi where $1_{{}_G}$ is the unit of $G$. 

In general a coproduct can be expanded on
$Fun(G)\otimes Fun(G)$ as:
\eq
\D(f)=\sum_i f_1^i \otimes f_2^i \equiv f_1 \otimes f_2, \label{not1}
\en
\noi where $f_1^i, f_2^i \in A=Fun(G)$ and $f_1 \otimes f_2$ is a 
shorthand notation we will often use in the sequel. Thus we have:
\eq
\D(f)(g,\gp)=(f_1 \otimes f_2)(g,\gp)=f_1(g)f_2(\gp)=f(g\gp). 
    \label{not2}
\en

It is not difficult to verify the following properties of the co-structures:
\begin{eqnarray}
 & & (id \otimes \D)\D=(\D\otimes id)\D~~~({\rm coassociativity~of~\D}) 
\label{prop1}\\
 & & (id\otimes \epsi)\D(a)=(\epsi\otimes id)\D(a)=a \label{prop2}\\
 & & m(\kappa\otimes id)\D(a)=m(id\otimes\kappa)\D(a)=\epsi(a) I 
\label{prop3}
\end{eqnarray}

\noi and
\begin{eqnarray}
 & & \D(ab)=\D(a)\D(b),~~~\D(I)=I\otimes I \label{prop4}\\
 & & \epsi(ab)~=\epsi(a) \epsi(b),~~~~~~\epsi(I)=1 \label{prop5} \\
 & & \kappa(ab) \; =\kappa(b)\kappa(a),~~~~~\kappa(I)=I \label{prop6}
\end{eqnarray}

\noi where $a,b \in A=Fun(G)$ and $m$ is the multiplication map
$m(a\otimes b)\equiv ab$. The product in $\D(a)\D(b)$ is the product in 
$A\otimes A$: $(a\otimes b)(c\otimes d)=ab\otimes cd$.

For example the coassociativity property $(\ref{prop1})$,
$(id \otimes \D)\D(f)=(\D\otimes id)\D(f)$ reads
$f_1\otimes (f_2)_{_1}\otimes (f_2)_{_2}=(f_1)_{_1}\otimes 
(f_1)_{_2}\otimes f_2$, for all $f \in A$.
This equality is easily seen to hold by applying it 
on the generic element $(g,g',g'')$ of 
$G\times G\times G$, and then by using associativity of the product in $G$.
\sk
An algebra $A$ (not necessarily commutative) endowed with the homomorphisms $\D:A \rightarrow A 
\otimes A$ and $\epsi: A \rightarrow \cc$, and the linear and 
antimultiplicative map   
$\kappa: A\rightarrow A$ satisfying the properties 
(\ref{prop1})-(\ref{prop6}) 
is a {\sl Hopf algebra}. Thus \Fun is a Hopf algebra,
it encodes the information on the group structure of $G$.

\sk

As a second example consider now the case where $G$ is a group of 
matrices, a subgroup of $GL$ given by matrices $\T{a}{b}$ that satisfy 
some algebraic relation (for example orthogonality conditions).
We then define $A=$\Fun to be the algebra of polynomials in the matrix
elements $\T{a}{b}$ of the defining representation of $G$ and in 
det$T^{-1}$; i.e. the algebra is generated 
by the $\T{a}{b}$ and det$T^{-1}$.
\sk
Using the elements $\T{a}{b}$ we can write an 
explicit formula for the expansion (\ref{not1}) or (\ref{not2}): indeed 
(\ref{cop}) becomes
\eq
\D(\T{a}{b})(g,\gp)=\T{a}{b} (g\gp)=\T{a}{c}(g) \T{c}{b}(\gp), 
\en
\noi since $T$ is a matrix representation of $G$. Therefore:
\eq
\D(\T{a}{b})=\T{a}{c} \otimes \T{c}{b}. \label{copT}
\en
Moreover, using (\ref{cou}) and (\ref{coi}), one finds:
\begin{eqnarray}
 & & \epsi(\T{a}{b})=\de^a_b \label{couT}\\
 & & \kappa(\T{a}{b})=\Ti{a}{b}. \label{coiT}
\end{eqnarray}
Thus the algebra $A=Fun(G)$ of polynomials in the elements $\T{a}{b}$ and 
det$T^{-1}$ is a Hopf algebra with co-structures defined by 
(\ref{copT})-(\ref{coiT}) and (\ref{prop4})-(\ref{prop6}).
\sk

The two example presented concern commutative Hopf algebras. 
In the first example the information on the group $G$ is equivalent to that on 
the Hopf algebra $A=$\Fun. We constructed $A$ from $G$. In order 
to recover $G$ from $A$ notice that every 
element $g\in G$ can be seen as a map 
from $A$ to $\cc$ defined by $f\rightarrow f(g)$. This map is multiplicative 
because $fh(g)=f(g)h(g)$. The set $G$ can be obtained 
from $A$ as the set of all nonzero multiplicative linear maps 
from $A$ to $\cc$ (the set of characters of $A$). 

Concerning the group structure of $G$, the product 
is recovered from the coproduct in $A$ via 
(\ref{not2}), i.e. $gg'$ is the new character that associates to any 
$f\in A$ the complex number $f_1(g)f_2(g')$. The unit of $G$ is the character 
$\epsi$; the inverse $g^{-1}$ is defined via the antipode of $A$.

In the second example, in order to recover the topology of $G$, 
we would need a $C^*$-algebra completion of the algebra 
$A=Fun(G)$ of polynomial functions. 
Up to these topological ($C^*$-algebra) aspects, we can say that the 
information concerning a matrix group $G$ can be encoded in its 
commutative Hopf algebra $A=Fun(G)$.
\sk
In the spirit of noncommutative geometry we now consider
noncommutative deformations $Fun_q(G)$ of the algebra $Fun(G)$. 
The space of points $G$ does not exist anymore, by
noncommutative or quantum space $G_q$ is meant the noncommutative algebra 
$Fun_q(G)$. Since $G$ is a group then $Fun(G)$ is a 
Hopf algebra; the noncommutative Hopf algebra obtained by deformation of 
$Fun(G)$ is then usually called
{\sl Quantum group}. The term quantum stems for the 
fact that the deformation
is obtained by quantizing a Poisson (symplectic) 
structure of the algebra $Fun(G)$ \cite{Drinfeld}.

\section{Quantum groups. The example of $SL_q(2)$}

Following \cite{FRT} we consider quantum 
groups defined as the associative algebras $A$ freely 
generated by non-commuting matrix entries 
$\T{a}{b}$ satisfying the relation
\eq
\R{ab}{ef} \T{e}{c} \T{f}{d} = \T{b}{f} \T{a}{e} \R{ef}{cd} \label{RTT}
\en
\noi and some other conditions depending on which classical group 
we are deforming (see later). The matrix $R$ controls the 
non-commutativity of the $\T{a}{b}$, and its elements depend 
continuously on a (in general complex) parameter $q$, or even a set of 
parameters. For $q\rightarrow 1$, the so-called ``classical limit", we
have 
\eq 
\R{ab}{cd} \qonelim \de^a_c \de^b_d,  \label{limR}
\en
\noi i.e. the matrix entries $\T{a}{b}$ commute for $q=1$, and one 
recovers the ordinary $Fun(G)$. 
The $R$-matrices for the quantum group deformation of the 
simple Lie groups of the $A,B,C,D$ series were given in \cite{FRT}.

The associativity of $A$ leads to a consistency condition on the $R$
matrix, the quantum Yang--Baxter equation:
\eq
\R{a_1b_1}{a_2b_2} \R{a_2c_1}{a_3c_2} \R{b_2c_2}{b_3c_3}=
\R{b_1c_1}{b_2c_2} \R{a_1c_2}{a_2c_3} \R{a_2b_2}{a_3b_3}. \label{YB}
\en
For simplicity we rewrite the ``RTT" equation (\ref{RTT}) and the 
quantum Yang--Baxter equation as
\eq
R_{12} T_1 T_2 = T_2 T_1 R_{12} \label{rtt}
\en
\eq
R_{12} R_{13} R_{23}=R_{23} R_{13} R_{12}, \label{yb}
\en
\noi where the subscripts 1, 2 and 3 refer to different couples of 
indices. Thus $T_1$ indicates the matrix $\T{a}{b}$, $T_1 T_1$ 
indicates $\T{a}{c} \T{c}{b}$, $R_{12} T_2$ indicates $\R{ab}{cd} 
\T{d}{e}$ and so on, repeated subscripts meaning matrix 
multiplication. The quantum Yang--Baxter equation (\ref{yb}) 
is a condition sufficient 
for the consistency of the RTT equation (\ref{rtt}). Indeed 
the product of three distinct elements $\T{a}{b}$, $\T{c}{d}$ 
and $\T{e}{f}$, indicated by $T_1T_2T_3$, can be reordered as 
$T_3T_2T_1$ via two differents paths:
\eq
 T_1T_2T_3 \begin{array}{c} \nearrow\\ \searrow \end{array}
           \begin{array}{c} T_1T_3T_2 \rightarrow T_3T_1T_2 \\ {} \\
                            T_2T_1T_3 \rightarrow T_2T_3T_1 \end{array}
           \begin{array}{c} \searrow\\ \nearrow \end{array}
 T_3T_2T_1          
\en
\noi by repeated use of the RTT equation. The relation (\ref{yb})
ensures that the two paths lead to the same result. 
\sk
The algebra $A$ (``the quantum group") is a noncommutative Hopf algebra 
whose co-structures are the same of those defined for the commutative 
Hopf algebra \Fun of the previous section, eqs. 
(\ref{copT})-(\ref{coiT}), (\ref{prop4})-(\ref{prop6}).
\sk
\noi{\bf Note} $~$ Define ${\hat R}^{ab}_{~~cd}=R^{ba}_{~~cd}$. Then the 
quantum Yang-Baxter equation becomes the braid relation
\eq
{\hat R}_{23}{\hat R}_{12}{\hat R}_{23}=
{\hat R}_{12}{\hat R}_{23}{\hat R}_{12}~.
\en 
If $\hat R$ satisfies ${\hat R}^2=id$ we have that $\hat R$ is 
a representation  of the permutation group. In the more general case
$\hat R$ is a representation of the braid group. The $\hat R$-matrx can be
used to construct invariants of knots.  
\sk
Let us give the example of the quantum group $SL_q(2)\equiv Fun_q(SL(2))$, the algebra freely 
generated by 
the elements $\al,\be,\ga$ and $\de$ of the $2 \times 2$ matrix
\eq
\T{a}{b}= \left( \begin{array}{cc} \al & \be \\ \ga & \de \end{array} 
    \right)   \label{Tmatrix}
\en
\noi satisfying the commutations
\eqa
\al\be=q\be\al,~~\al\ga=q\ga\al,~~\be\de=q\de\be,~~\ga\de=q\de\ga
\nonumber \\
\be\ga=\ga\be,~~\al\de-\de\al=(q-\qm)\be\ga,~~~~q\in\cc
\label{sucomm} 
\ena
\noi and
\eq
{\det}_q T\equiv\al\de-q\be\ga=I. \label{sudet}
\en
The commutations (\ref{sucomm}) can be obtained from (\ref{RTT}) via the 
$R$ matrix
\eq
\R{ab}{cd}=\left( \begin{array}{cccc} q & 0 & 0 & 0 \\
                                      0 & 1 & 0 & 0 \\
                                      0 & q-\qm & 1 & 0 \\
                                      0 & 0 & 0 & q  \end{array} \right)
  \label{Rsu}
\en
\noi where the rows and columns are numbered in the order 
11, 12, 21, 22. 

It is easy to verify that the ``quantum determinant" defined in 
(\ref{sudet}) commutes with $\al,\be,\ga$ and $\de$, so that the 
requirement ${\det}_q T=I$ is consistent. The matrix inverse of $\T{a}{b}$
is 
\eq
\Ti{a}{b}= ({\det}_q T)^{-1} \left( \begin{array}{cc} \de & -\qm\be \\ 
          -q\ga & \al 
         \end{array} \right)   \label{Timatrix}~.
\en
\sk
The coproduct, counit and coinverse of $\al,\be,\ga$ and $\de$ are 
determined via formulas (\ref{copT})-(\ref{coiT}) to be:
\eqa
\D(\al)=\al\otimes\al+\be\otimes\ga,~~~\D(\be)=\al\otimes\be+
\be\otimes\de \nonumber\\
\D(\ga)=\ga\otimes\al+\de\otimes\ga,~~~\D(\de)=\ga\otimes\be+
\de\otimes\de \label{copsu}\\
\epsi(\al)=\epsi(\de)=1,~~~\epsi(\be)=\epsi(\ga)=0 \label{cousu}\\
\kappa(\al)=\de,~~\kappa(\be)=-\qm \be,~~\kappa(\ga)=-q\ga,~~
\kappa(\de)=\al . \label{coisu}
\ena
\sk
\noi{\bf Note} $~$ The commutations (\ref{sucomm}) are compatible with 
the 
coproduct $\D$, in the sense that $\D(\al\be)=q\D(\be\al)$ and so on.
In general we must have 
\eq
\D(R_{12}T_1T_2)=\D(T_2T_1R_{12}), \label{DRTT} 
\en
\noi which is easily verified using $\D(R_{12}T_1T_2)=R_{12}\D(T_1)
\D(T_2)$ and $\D(T_1)=T_1 \otimes T_1$. This is equivalent to proving 
that the matrix elements of the matrix product $T_1\TP_1$, where
$\TP$ is a matrix [satisfying (\ref{RTT})] 
whose elements {\sl commute} with those of $\T{a}{b}$,
still obey the commutations (\ref{rtt}).
\sk
\noi{\bf Note} $~\D({\det}_qT)={\det}_qT\otimes {\det}_qT~$ so that the coproduct 
property $\D(I)=I\otimes I$ is compatible with ${\det}_qT=I$.
\sk
\noi{\bf Note} $~$ The condition (\ref{sudet}) can be relaxed. Then we have 
to include the central element $\zeta=({\det}_q T)^{-1}$ in $A$, so 
as to be able 
to define the inverse of the $q$-matrix $\T{a}{b}$ 
as in (\ref{Timatrix}),
and the coinverse of the element $\T{a}{b}$ as in (\ref{coiT}). The
$q$-group is then $GL_q(2)$. The reader can deduce the co-structures on 
$\zeta$: $\D(\zeta)
=\zeta \otimes \zeta,~\epsi(\zeta)=1,~\kappa(\zeta)={\det}_q T$.
\sk
\section{Universal enveloping algebras and 
$\U$}
Another example of Hopf algebra is given by any ordinary Lie algebra 
$\mbox{\sl g}$, or more precisely by the universal enveloping algebra 
$U(\mbox{\sl g})$ of a Lie algebra $\mbox{\sl g}$, i.e. 
the 
algebra, with unit $I$, of polynomials in the generators $\chi_i$ modulo 
the commutation relations
\eq
[\chi_i,\chi_j]=\c{ij}{k} \chi_k~. \label{clcomm}
\en
Here we define the co-structures on the generators as:
\begin{eqnarray}
 & & \D(\chi_i)=\chi_i \otimes I + I \otimes \chi_i~~~\D(I)=I\otimes I  
  \label{copL}\\
 & & \epsi (\chi_i)=0~~~~~~~~~~~~~~~~~~~~~~\epsi (I)=1 \label{couL}\\
 & & \kappa(\chi_i)=-\chi_i~~~~~~~~~~~~~~~~~~\kappa (I)=I \label{coiL}
\end{eqnarray}
and extend them to all $U(\mbox{\sl g})$ by requiring 
$\D$ and $\epsi$ to be linear and multiplicative, 
$\D(\chi\chi')=\D(\chi)\D(\chi')$, 
$\epsi(\chi\chi') =\epsi(\chi)\epsi(\chi')$, while $S$ is linear and antimultiplicative. In order to show that the construction of the Hopf algebra 
$U(\mbox{\sl g})$ is well given, we have to check that the maps $\D,\epsi,S$ are well defined. We give the proof for the coproduct. 
Since $[\chi,\chi']$ is linear in the generators we have
\eq
\D[\chi,\chi']=[\chi,\chi']\otimes I +I\otimes [\chi,\chi']~,
\en 
on the other hand, using that $\D$ is multiplicative we have
\eq
\D[\chi,\chi']=\D(\chi)\D(\chi')-\D(\chi')\D(\chi)
\en
it is easy to see that these two expressions coincide.

The Hopf algebra $U(\mbox{\sl g})$ is noncommutative but it is cocommutative,
i.e. for all $\zeta\in U(\mbox{\sl g})$, 
$\zeta_1\otimes\zeta_2=\zeta_2\otimes\zeta_1$, 
where we used the notation $\D(\zeta)=\zeta_1\otimes\zeta_2$.
We have discussed deformations of commutative Hopf algebras, of the kind 
$A=Fun(G)$, 
and we will see that these are related to deformations of 
cocommutatative Hopf algebras of the kind $U(g)$ where 
$g$ is the Lie algebra of $G$.
\sk
We here introduce the basic 
example of deformed universal enveloping algebra: 
$\U$ \cite{Drinfeld, Jimbo}, which is a deformation of the usual enveloping algebra of $sl(2)$,
\eq
[X^+,X^-]=H~~,~~~[H,X^{\pm}]=2X^\pm~.
\en
The Hopf algebra $\U$ is generated by the elements $K_+$, $K_-$,
$X_+$ and $X_-$ and the unit element $I$, that satisfy the relations   
\eqa
& & [X_+,X_-]={{K_+^2-K_-^{-2}}\over {q-q^{-1}}}~,\label{sl2uno}\\
& & K_+ X_{\pm}K_-=q^{\pm 1}X_\pm~,\label{sl2due}\\
& & K_+ K_-=K_- K_+=I~.
\ena
The parameter $q$ that appears 
in the right hand side of the first two equations is a complex number.
It can be checked that the algebra $\U$ becomes a Hopf algebra by defining 
the following costructures
\begin{eqnarray}
 & & \D(X_\pm)=X^\pm\otimes K_+ + K_- \otimes X_\pm~,~~
\D(K_\pm)=K_\pm\otimes K_\pm   
  \label{cosl2}\\
 & & \epsi (X_\pm)=0~~~~~~~~~~~~~~~~~~~~~~~~~~~~~~~~~\epsi (K_\pm)=1 \label{cousl2}\\
 & & \kappa(X_\pm)=-q^{\pm 1}X_\pm~~~~~~~~~~~~~~~~~~~~~~~
\kappa(K_\pm)=K_\mp \label{colis2}
\end{eqnarray}
If we define $K_+=1+{q\over 2}{\cal H}$ then we see that in the limit $q\rightarrow 1$
we recover the undeformed $U(sl(2))$ Hopf algebra. 

The Hopf algebra $\U$ is not cocommutative, however the noncocommutativity 
is under control, as we now show.
We set $q=e^h$, consider $h$ a formal parameter and allow for power series in $h$.
We are considering a topological completion of $\U$, this is equivalently 
generated by the three generators $X_\pm$ and $H$, where we have 
$K_\pm=e^{\pm h H/2}$. In this case there exists  an element $\cal R$ 
of $\U\widehat{\otimes}\U$ (also the usual tensorproduct $\otimes$ has to 
be extended to allow for power series), called universal $R$-matrix that 
governs the noncocommutativity of the coproduct $\D$,
\eq
\sigma\D(\zeta)={\cal R}\D(\zeta){\cal R}^{-1}~,
\en
where $\sigma$ is the flip operation, $\sigma(\zeta\otimes\xi)=\xi\otimes\zeta$.
The element ${\cal R}$ explicitly reads
\eq
{\cal R}=q^{{H\otimes H}\over 2} \sum_{n=0}^\infty{{(1-q^{-2})^n}\over {[n]!}}
(q^{H/2} X_+\otimes q^{-H/2}X_-)^n q^{n(n-1)/2}
\en
where $[n]\equiv {{q^n-q^{-n}}\over {q-q^{-1}}}$, and $[n]!=[n][n-1]\ldots 1$.
\sk
The universal $\cal R$ matrix has further properties, that strucure $\U$ to be a
quasitriangular Hopf algebra. Among these properties we mention that ${\cal R}$
is invertible and that it satifies the Yang-Baxter equation
\eq
{\cal R}_{12}{\cal R}_{13}{\cal R}_{23}={\cal R}_{23}{\cal R}_{13}{\cal R}_{12}
\en
where we used the notation ${\cal R}_{12}={\cal R}\otimes I$,  
${\cal R}_{23}=I\otimes {\cal R}$ and 
${\cal R}_{13}={\cal R^\al}\otimes I\otimes {\cal R}_\al$, where
${\cal R}={\cal R}^\al\otimes {\cal R}_\al$ (sum over $\al$ understood).
\sk

\section{Duality
}

\sk
Consider a finite dimensional Hopf algebra $A$, the vector 
space $A'$ dual to $A$ is also a Hopf algebra with the following  
product, unit and costructures
[we use the notation $\psi(a)=\le\psi,a\re$ in order to stress the
duality between $A'$ and $A$]: $\forall\psi,\phi\in A'$, $\forall a,b\in A$
\eq
\le \psi\phi,a\re=\le\psi\otimes\phi,\D a\re~,~~\le I,a\re=\epsi(a)
\label{pair1}
\en
\eq
\le\D(\psi), a\otimes b\re=\le\psi,ab\re~,~~\epsi(\psi)=\le\psi,I\re
\label{pair2}
\en
\eq \le \kappa (\psi),a\re=\le\psi,\kappa (a)\re\label{pair3}
\en 
where 
$\le \psi\otimes \phi~,~a\otimes b\re\equiv\le\psi, a\re\,\le\phi, b\re~.$ 
Obviously $(A')'=A$ and $A$ and $A'$ are dual Hopf algebras.
\sk
In the infinite dimensional case the definition of duality between 
Hopf algebras is more delicate because the coproduct on $A'$ 
might not take values  in the subspace  $A'\otimes A'$ of
$(A\otimes A)'$ and therefore is ill defined.
We therefore use the notion of pairing:
two Hopf algebras $A$ and $U$ are paired if there exists a bilinear map
$\le~,~\re~:U\otimes A\rightarrow \cc$ satisfying (\ref{pair1}) and 
(\ref{pair2}), (then (\ref{pair3}) can be shown to follow as well).
%
\sk
The Hopf algebras $Fun(G)$ and $U(\mbox{\sl g})$ 
described in Section 3 and 4 are paired if $\mbox{\sl g}$ is 
the Lie algebra of $G$.
Indeed we realize $\mbox{\sl g}$ as left invariant vectorfields on the 
group manifold.
Then the pairing is defined by
\[\forall t\in \mbox{\sl g}, \forall f\in Fun(G),~~
\le t,f\re=t(f)|_{{}_{1_{{}_G}}}~,
\] where $1_{{}_G}$ is the unit of $G$, and more in general is well defined 
by\footnote{In order to see that relations (\ref{pair1}),(\ref{pair2})
hold, we recall that        
$t$ is left invariant if $TL_g(t|_{{}_{1_{{}_G}}})=
t|_g$, where $TL_g$ is the tangent map induced by the left multiplication
of the group on itself: $L_gg'=gg'$.  We then have  
$$
t(f)|_g=\left( TL_g t|_{{}{1_{{}_G}}}\right)(f)=
t[f(g{\tilde g})]|_{{}_{{\tilde g}=1_{{}_G}}}=  
t[f_1(g)f_2({ \tilde g})]|_{{}_{{\tilde g}=1_{{}_G}}}=
f_1(g)\,t(f_2)|_{{}_{1_{{}_G}}}\label{tislinv}
$$
and therefore
\[\le \tilde{t}\, t ,f\re=\tilde{t} 
(t(f))|_{{}_{1_{{}_G}}}=
\tilde{t}f_1|_{{}_{1_{{}_G}}}tf_2|_{{}_{1_{{}_G}}}
=\le \tilde{t}\otimes t,\D f\re~,
\]
and 
\[\le t, f h\re=t(f)|_{{}_{1_{{}_G}}}h|_{{}_{1_{{}_G}}}+f|_{{}_{1_{{}_G}}}
t(h)|_{{}_{1_{{}_G}}}=\le \D(t),f\otimes h\re~.
\]}
\[\forall\, tt'...t''\in U(\mbox{\sl g}), \forall f\in Fun(G),~~
\le tt'...t'',f\re=t(t'...(t''(f)))|_{{}_{1_{{}_G}}}~.
\] 
\sk

The duality between the Hopf algebras $Fun(Sl(2))$ and $U(sl(2))$ holds also 
in the deformed case, so that the quantum group $Sl_q(2)$ is dual to $\U$.
In order to show this duality we introduce a subalgebra
(with generators $L^\pm$) of the algebra of linear maps from 
$Fun_q(Sl(2))$ to $\cc$. We then see that this subalgebra has a natural Hopf
algebra structure dual to $SL_q(2)=Fun_q(Sl(2))$. Finally we see in formula
(\ref{usl2ll}) that this subalgebra is just $\U$.
This duality is important because it allows to consider the elements of 
$\U$ as (left invariant) differential operators on $SL_2(2)$. 
This is the first step 
for the construction of a differential calculus on the quantum group 
$Sl_q(2)$. 
\sk
\noi{\bf The $L^{\pm}$ functionals}

\noi The linear functionals $\Lpm{a}{b}$ are defined
by their value on the elements $\T{a}{b}$:
\eq
\Lpm{a}{b} (\T{c}{d})=\le\Lpm{a}{b} , \T{c}{d}\re=R^\pm{}^{ac}_{~~bd}, 
\label{defL}
\en
\noi where
\eq
(R^+)^{ac}_{~~bd} \equiv  q^{-1/2}R^{ca}_{~~db} \label{Rplus} 
\en
\eq
(R^-)^{ac}_{~~bd} \equiv q^{1/2} (R^{-1})^{ac}_{~~bd}, \label{Rminus}
\en
The inverse matrix $R^{-1}$ is defined by
\eq
{R^{-1}}^{ab}_{~~cd}\R{cd}{ef} \equiv \de^
a_e \de^b_f \equiv \R{ab}{cd} {R^{-1}}^{cd}_{~~ef}. 
\en
To extend the definition (\ref{defL}) 
to the whole algebra $A$ we set:
\eq
\Lpm{a}{b} (ab)=\Lpm{a}{g} (a) \Lpm{g}{b} (b),~~~\forall a,b\in A 
\label{Lab}
\en
\noi so that, for example,
\eq
\Lpm{a}{b} (\T{c}{d} \T{e}{f}) = {R^\pm}^{ac}_{~~gd}{R^\pm}^{ge}_{~~bf}.\label{perQYB}
\en
\noi In general, using the compact notation introduced in Section 2,
\eq
\LLpm_1(T_2T_3...T_n)=R^\pm_{12} R^\pm_{13} ... R^\pm_{1n}. \label{LTT}  
\en
\noi As it is esily seen from (\ref{perQYB}), the quantum Yang-Baxter equation 
(\ref{yb}) is a {necessary} and {sufficient} condition for the 
compatibility of (\ref{defL}) and (\ref{Lab}) with the $RTT$ relations:
$\LLpm_1(R_{23}T_2T_3 - T_3T_2R_{23})=0$.
\sk
Finally, the value of $\LLpm$ on the unit $I$ is defined by
\eq
\Lpm{a}{b} (I)=\de^a_b. \label{LI}
\en
It is not difficult to find the commutations between $\Lpm{a}{b}$ 
and $\Lpm{c}{d}$:
\eq
R_{12} \LLpm_2 \LLpm_1=\LLpm_1 \LLpm_2 R_{12} \label{RLL}
\en
\eq 
R_{12} \LLp_2 \LLm_1=\LLm_1 \LLp_2 R_{12}, \label{RLpLm}
\en
\noi where the product $\LLpm_2 \LLpm_1$ is by definition obtained by duality from the coproduct in $A$, for all $a\in A$, $$\LLpm_2 \LLpm_1 (a)\equiv (\LLpm_2 \otimes \LLpm_1)\D(a)~.$$ 
For example consider
\eq
R_{12} (\LLp_2 \LLp_1)(T_3)=R_{12}(\LLp_2 \otimes\LLp_1)\D (T_3)=
R_{12}(\LLp_2 \otimes\LLp_1)(T_3 \otimes T_3)=q\:R_{12}R_{32}R_{31}
\nonumber
\en
\noi and
\eq
\LLp_1 \LLp_2(T_3) R_{12}=q \: R_{31}R_{32}R_{12}
\nonumber
\en
\noi so that the equation (\ref{RLL}) is proven for $\LLp$ 
by virtue of the 
quantum Yang--Baxter equation (\ref{YB}), where the indices have been 
renamed $2\rightarrow 1,3 \rightarrow 2,1\rightarrow 3$. Similarly,
one proves the remaining ``RLL" relations.
\sk
\noi{\bf Note} $~$ As mentioned in \cite{FRT}, $L^+$ is upper 
triangular, $L^-$ is lower triangular (this is due to the upper 
and lower triangularity of $R^+$ and $R^-$, respectively).
{}From (\ref{RLL}) and (\ref{RLpLm}) we have
\eq
\Lpm{a}{a} \Lpm{b}{b}=\Lpm{b}{b} \Lpm{b}{b}~;~~ 
\Lp{a}{a} \Lm{b}{b}=\Lm{b}{b} \Lp{a}{a}=\epsi ~.\label{LLrelations}
\en
we also have
\eq
\Lpm{1}{1} \Lpm{2}{2}=\epsi~.\label{detrelation}
\en
\sk 
\sk
The algebra of polynomials in the $L^\pm$ functionals
becomes a Hopf algebra paired to $Sl_q(2)$ by defining 
the costructures via the duality (\ref{defL}):
\eq
\D(\Lpm{a}{b})(\T{c}{d} \otimes \T{e}{f}) \equiv \Lpm{a}{b}
(\T{c}{d}\T{e}{f})=\Lpm{a}{g}(\T{c}{d}) \Lpm{g}{b} (\T{e}{f})
\en
\eqa
& & \epsi (\Lpm{a}{b})\equiv \Lpm{a}{b} (I)\\[1.3em]
& & S (\Lpm{a}{b})(\T{c}{d})\equiv \Lpm{a}{b} (\kappa (\T{c}{d})) 
\ena
\noi cf. [(\ref{pair2}), (\ref{pair3})], so that
\eqa
& & \D (\Lpm{a}{b})=\Lpm{a}{g} \otimes \Lpm{g}{b}\\
& & \epsi (\Lpm{a}{b})=\de^a_b \\
& & S (\Lpm{a}{b})=\Lpm{a}{b} \circ \kappa ~.
\ena
\noi
This Hopf algebra is $\U$ because it can be checked that relations 
(\ref{RLL}), (\ref{RLpLm}), (\ref{LLrelations}), (\ref{detrelation}) 
fully characterize the $L^\pm$ functionals, so that the algebra of polynomials 
in the {\sl symbols} ${L^\pm}^a_{~b}$ that satisfy the relations   
(\ref{RLL}), (\ref{RLpLm}), (\ref{LLrelations}), (\ref{detrelation}) is 
isomorphic to the algebra generated by the $L^\pm$ {\sl functionals} on $\U$. 
An explicit
relation between the $L^\pm$ matrices and the generators $X^\pm$ and $K^\pm$
of $\U$ introduced in the previous section is obtained by comparing the ``RLL''
commutation relations with the $\U$ Lie algebra relations, we obtain

\eq\label{usl2ll}
L^+=\left( \begin{array}{cc} K_- & ~~q^{-1/2}(q-q^{-1})X_+  \\ 0 & K_+ 
\end{array} 
\right)~~~,~~~~~
L^-=\left( \begin{array}{cc} K_+ & \,0  \\ q^{1/2}(q^{-1}-q)X_- ~~& K_- 
\end{array} 
\right)~.
\en
\sk
\section{Quantum Lie algebra}  
\sk
We now turn our attention to the issue of determining the Lie algebra 
of the quantum group $Sl_q(2)$, or equivalently the quantum
Lie algebra of the universal enveloping algebra $\U$.

In the undeformed case the Lie algebra of a universal enveloping algebra 
$U$ (for example $U(sl(2))$) is the unique linear 
subspace $g$ of $U$ of primitive elements, 
i.e. of elements $\chi$ 
that have coproduct: 
\eq
\Delta(\chi)=\chi\otimes 1 +1\otimes \chi~ .
\en
Of course $g$ generates $U$ and $g$ is closed under the usual 
commutator bracket $[~,~]$,
\eq
[u,v]= u u-v u\in g ~~~~\mbox{for all } u, v\in g~.
\en
The geometric meaning of the bracket  $[u,v]$ is that it is the 
adjoint action of $g$ on $g$,
\eq\label{adactcomm}
[u,v]=ad_u\,v
\en
\eq\label{edfr}
ad_u\,v:=u_1vS(u_2)
\en
where we have used the notation $\D(u)=\sum_\al u_{1_\al}\otimes u_{2_\al}=u_1\otimes u_2$, 
so that a sum over $\al$ is understood. Recalling that 
$\D(u)=u\otimes 1+1\otimes u$ and that $S(u)=-u$, from (\ref{edfr})
we immediately obtain
(\ref{adactcomm}). In other words, the commutator $[u,v]$ 
is the Lie derivative of the left invariant vectorfield  
$u$ on the left invariant vectorfield $v$. 
More in general the adjoint action of $U$ on  
$U$ is given by 
\eq
ad_\xi\,\zeta=\xi_1\zeta S(\xi_2)~,
\en
where we used the notation (sum understood)
$\D(\xi)=\xi_1\otimes\xi_2~.$
\sk
In the deformed case the coproduct is no more cocommutative and  
we cannot identify the Lie algebra of a deformed universal enveloping algebra $U_q$  with the primitive elements
of $U_q$, they are too few to generate $U_q$. We then have to relax this requirement.   
There are three natural conditions that according to
\cite{Woronowicz}
the $q$-Lie algebra of a $q$-deformed universal enveloping algebra 
$U_q$ has to satisfy, see \cite{AC,SWZ}, and \cite{{AschieriTesi}} p. 41. 
It has to be a linear subspace $g_q$ of $U_q$ such that
\eqa
i)&&g_q \mbox{ generates } U_q~, \\
ii)&&\D(g_q)\subset g_q\otimes 1+\U_q\otimes g_q ~,\\
iii)&&[g_q,g_q]\subset g_q~.
\ena
Here now $\D$ is the coproduct of $U_q$ and $[~,~]$ denotes the
$q$-bracket 
\eq
[u,v]=ad_u\,v=u_{1} v S(u_{2})~.
\en
where we have used the coproduct notation $\D(u)=u_{1}\otimes u_{2}$. 
Property $iii)$ is the closure of $g_q$ under the adjoint action.
Property $ii)$ implies a minimal deformation of the Leibnitz rule. 

From these conditions, that do not in general single out a 
unique subspace $g_q$, it follows that the bracket $[u,v]$ 
is quadratic in $u$ and $v$, that it has a deformed 
antisymmetry property and that it satisfies a 
deformed Jacoby identity. 
\sk
In the example $U_q=\U$ we have that a quantum $sl(2)$ Lie algebra is  
spanned by the four linearly independent elements 
\eq
\chi^{c_1}_{~~c_2}={1\over{q-q^{-1}}}[\Lp{c_1}{b} S(\Lm{b}{c_2})
-\de^{c_1}_{c_2} \epsi ]~.
\label{defchi2}
\en
In the commutative limit $q\rightarrow 1$, we have 
$\chi^2_{~2}=-\chi^1_{~1}=H/2$, $\chi^1_{~2}=X_+$, $\chi^2_{~1}=X_-$, 
and we recover the usual $sl(2)$ Lie algebra.
\sk
The $q$-Lie algebra commutation relations explicitly are
\[ \chi_1 \chi_+ - \chi_+ \chi_1 +(q^4-q^2)\chi_+ \chi_2 = q^3 \chi_+\]
\[ \chi_1 \chi_- - \chi_- \chi_1 -(q^2-1)\chi_- \chi_-2= -q \chi_-\]
\[ \chi_1 \chi_2 - \chi_2 \chi_1 =0\]
\[ \chi_+ \chi_- - \chi_- \chi_+ - (1-q^2) \chi_1 \chi_2+(1-q^2) 
\chi_2 \chi_2 = q (\chi_1-\chi_2)
\]
\[ \chi_2 \chi_+-q^2 \chi_+ \chi_2 = -q\chi_+\]
\[ \chi_2 \chi_- -q^{-2} \chi_- \chi_2 = \qm \chi_-\]
where we used the composite index notation
$$
{}^{a_1}_{~~a_2}\rightarrow \,_i~,~~_{b_1}^{~~b_2}\rightarrow\, ^j~,
~~{\mbox{ and }}~~ i,j=1,+,-,2~.
$$
These $q$-lie algebra relations can be compactly 
written \cite{AC}\footnote{Relation to the conventions of \cite{Woronowicz, AC}
(here underlined): $\chi_i=-S^{-1}\underline{\chi}_i$, 
$f_j^{~i}=S^{-1}\underline{f}^i_{~j}$.}
\eq
[\chi_i,\chi_j]=\chi_i\chi_j-\La^{rs}_{~~ji}\chi_s\chi_r~,
\en
where 
$\La_{a_1}^{~a_2} {}_{d_1}^{~d_2}{\:}^{c_1}_{~c_2} {}^{b_1}_{~b_2}=
S(\Lp{b_1}{a_1}) \Lm{a_2}{b_2} (\T{c_1}{d_1} 
S(\T{d_2}{c_2}))$.
The $q$-Jacoby identities then read
\eq
[\chi_i,[\chi_j,\chi_r]]=[[\chi_i,\chi_j],\chi_r]]+\La^{kl}_{~~ji}[\chi_l,[\chi_k,\chi_r]]~.
\en

\section{Deformation by twist and quantum Poincar\'e Lie algebra}
In this last section, led by the example the Poincar\'e Lie algebra, we 
rewiew a quite general method to deform the Hopf algebra $U(g)$, the 
universal enveloping algebra of a given Lie algebra $g$. It is based 
on a twist procedure.
A twist element $\FF$ is an invertible element in $U(g)\otimes U(g)$.
A main property $\FF$ has to satisfy is the cocycle condition
\eq
\label{propF1}
(\FF\otimes 1)(\Delta\otimes id)\FF=(1\otimes \FF)(id\otimes \Delta)\FF\,~.
\en
Consider for example the usual Poincar\'e Lie algebra $iso(3,1)$:
\eqa\label{LiePoinc} [P_\mu,P_\nu]&=&0\ ,\nn\\[.3em]
[P_\rho , M_{\mu\nu}]&=&i(\eta_{\rho\mu}P_\nu-\eta_{\rho\nu}P_\mu)\ ,\\[.3em]
[M_{\mu\nu},M_{\rho\sigma}]&=&-i(\eta_{\mu\rho}
M_{\nu\sigma}-\eta_{\mu\sigma}M_{\nu\rho}
-\eta_{\nu\rho}M_{\mu\sigma}+\eta_{\nu\sigma}M_{\mu\rho})\ ,
\ena
A twist element is given by 
\eq
\FF=e^{{i\over 2}\theta^{\mu\nu}{P_\mu}\otimes{P_\nu}}~,
\en
where $\theta^{\mu\nu}$ (despite the indices $\mu\nu$ notation)
is a real antisymmetric matrix of 
dimensionful constants (the previous deformation parameter $q$ was a consant too). 
We consider $\theta^{\mu\nu}$  fundamental physical constants,
like the velocity of light $c$, or like $\hbar$. In this setting 
symmetries will leave $\theta^{\mu\nu}, \,c$ and $\hbar$ invariant.
The inverse of $\FF$ is 
$$\FF=e^{{-i\over 2}\theta^{\mu\nu}{P_\mu}\otimes{P_\nu}}~.$$
This twist satisfies the cocycle condition (\ref{propF1}) 
because the Lie algebra of momenta is abelian.

The Poincar\'e Hopf algebra $U^\FF(iso(3,1))$ is a deformation of $\UP$.
As algebras 
$U^\FF(iso(3,1))=U(iso(3,1))$; but $U^\FF(iso(3,1))$ has the new coproduct
\eq\label{twistedcoprodct}
\D^\FF(\xi)=\FF\D(\xi)\FF^{-1} ~,
\en
for all $\xi\in U(iso(3,1))$. 
In order to write the explicit expression for  $\D^\FF(P_\mu)$ and 
$\D^\FF(M_{\mu\nu})$, we use the Hadamard formula
$$Ad_{e^X}Y=e^X\ Y\
e^{-X}=\sum\limits_{n=0}^\infty\frac{1}{n!}\underbrace{[X,[X,...[}_n
X,Y]]=\sum\limits_{n=0}^\infty\frac{(ad X)^n}{n!}\ Y$$ and 
the relation $[P\otimes P', M\otimes 1]=[P,M]\otimes P'$, and thus obtain 
\cite{Chaichian}, \cite{Wess} 
\eqa
\Delta^\FF(P_\mu)&=&P_\mu\otimes1+1\otimes
P_\mu~,\nn\\[.3em]
\label{copPoin} 
\Delta^\FF(M_{\mu\nu})&=&M_{\mu\nu}\otimes 1+1 \otimes
M_{\mu\nu}\\& & ~-\frac{1}{2}\theta^{\alpha\beta}\left((\eta_{\alpha\mu}
P_\nu-\eta_{\alpha\nu}P_\mu)\otimes P_\beta
+P_\alpha\otimes(\eta_{\beta\mu}P_\nu-\eta_{\beta\nu}P_\mu)\right)
.\nn\ena
We have constructed the Hopf algebra $U^\FF(iso(3,1)$: it is the algebra 
generated by $M_{\mu\nu}$ and $P_\mu$ modulo the relations (\ref{LiePoinc}),
and with coproduct (\ref{copPoin}) and counit and 
antipode that are as in the undeformed case:
\eq
\epsi(P_\mu)=\epsi(M_{\mu\nu})=0~~,~~~
S(P_\mu)=-P_\mu~~,~~~S(M_{\mu\nu})=-M_{\mu\nu}~.
\en
This algebra is a symmetry algebra of the noncommutative spacetime 
$x^\mu x^\nu-x^\nu x^\mu=i\theta^{\mu\nu}$.
\sk
In general given a Lie algebra $g$, and a twist $\FF\in U(g)\otimes U(g)$,
formula (\ref{twistedcoprodct}) defines a new coproduct that is not
cocommutative. We call $U(g)^\FF$ the new Hopf algebra with coproduct $\D^\FF$,
counit $\epsi^\FF=\epsi$ and antipode $S^\FF$ that is a deformation of 
$S$ \cite{Drinfeldtwist}\footnote{Explicitly, if we write 
$\FF={\rm f}\,^\al\otimes {\rm f}\,_\al$ and define 
the element $\chi={\rm f}^\al S({\rm f}_\al)$ 
(that can be proven to be invertible) then for all elements $\xi\in U(g)$,
$S^\FF(\xi)=\chi S(\xi)\chi^{-1}$.}. 
By definition as algebra $U(g)^\FF$ equals $U(g)$, only the costructures 
are deformed.
\sk
We now construct the quantum Poincar\'e Lie algbra $iso^{\FF\!}(3,1)$.
Following the previous section, the Poincar\'e Lie algebra $iso^{\FF\!}(3,1)$
must be a linear subspace of $U^\FF(iso(3,1))$ such that
if $\{t_i\}_{i=1,...n}$ is a basis of $iso^{\FF\!}(3,1)$, we have (sum understood on repeated indices)
\eqa
i)&&\{t_i\} \mbox{ generates } U^\FF(iso(3,1)) \nn\\[.2em]
ii)&&\D^\FF(t_i)= t_i\otimes 1+f_i{}^j\otimes t_j \nn\\[.2em]
iii)&&[t_i,t_j]_\FF=C_{ij}{}^kt_k\nn
\ena
where $C_{ij}{}^k$ are structure constants and $f_i{}^j\in U^\FF(iso(3,1))$ 
($i,j=1,...n$).
%
In the last line the bracket $[~,~]_{_\FF}$ is the adjoint action:
\eq
[t,t']_{_\FF}:=ad^\FF_t\,t'=t_{1_\FF}t'S(t_{2_\FF})~,
\en
where we used the coproduct notation $\D^\FF(t)=t_{1_\FF}\otimes t_{2_\FF}$.
The statement that the {Lie algebra} of 
$U^\FF(iso(3,1))$ is the undeformed Poincar\'e Lie algebra (\ref{LiePoinc})
is not correct because conditions $ii)$
and $iii)$ are not met by the generators $P_\mu$ and $M_{\mu\nu}$.
There is a canonical procedure in order to obtain the Lie algebra 
$iso^{\FF\!}(3,1)$ of 
$U^\FF(iso(3,1))$ \cite{GR2, Aschieri2006}. Consider the elements
\eqa
P_\mu^\FF&:=&\of^\al(P_\mu)\of_\al=P_\mu~,\\[.3em]
M_{\mu\nu}^\FF&:=&\of^\al(M_{\mu\nu})\of_\al 
=M_{\mu\nu}-{i\over 2}
\theta^{\rho\sigma}[P_\rho,M_{\mu\nu}]P_\sigma
\nn\\&~&~~~~~~~~~~~~~~~=M_{\mu\nu}+{1\over 2}
\theta^{\rho\sigma}(\eta_{\mu\rho}P_\nu-
\eta_{\nu\rho}P_{\mu})P_\sigma
\ena
Their coproduct is 
\eqa
\Delta^\FF(P_\mu)&=&P_\mu\otimes1+1\otimes
P_\mu~,\nn\\[.3em]
\label{copPoinF} 
\Delta^\FF(M^\FF_{\mu\nu})&=&
M^\FF_{\mu\nu}\otimes 1+1 \otimes
M^\FF_{\mu\nu}+i\theta^{\alpha\beta}P_\al\otimes [P_\be,M_{\mu\nu}]~.
\ena
The counit and antipode are
\eq
\epsi(P_\mu)=\epsi(M^\FF_{\mu\nu})=0~~,~~~
S(P_\mu)=-P_\mu~~,~~~S(M^\FF_{\mu\nu})=-M^\FF_{\mu\nu}
-i\theta^{\rho\sigma}[P_\rho,M_{\mu\nu}]P_\sigma~.
\en
The elements $P^\FF_\mu$ and $M^\FF_{\mu\nu}$ are generators because they 
satisfy condition $i)$ (indeed $M_{\mu\nu}=M^\FF_{\mu\nu}+{i\over 2}
\theta^{\rho\sigma}[P_\rho,M^\FF_{\mu\nu}]P_\sigma$). They are deformed 
{\sl infinitesimal} generators because they satisfy the Leibniz rule $ii)$ 
and because they close under the Lie bracket $iii)$. Explicitly 
\eqa\label{LiePoincF}
 [P_\mu,P_\nu]_{_\FF}&=&0\ ,\nn\\[.3em]
[P_\rho , M^\FF_{\mu\nu}]_{_\FF}&=&i(\eta_{\rho\mu}P_\nu-\eta_{\rho\nu}P_\mu)
\,,\nn\\[.3em]
[M^\FF_{\mu\nu},M^\FF_{\rho\sigma}]_{_\FF}&=&
-i(\eta_{\mu\rho}M^\FF_{\nu\sigma}-\eta_{\mu\sigma}M^\FF_{\nu\rho}
-\eta_{\nu\rho}M^\FF_{\mu\sigma}+\eta_{\nu\sigma}M^\FF_{\mu\rho})~.
\ena
We notice that the structure constants are the same as in the undeformed case,
however the adjoint action $[M^\FF_{\mu\nu},M^\FF_{\rho\sigma}]_{_\FF}$ 
is not the commutator anymore, it is a deformed commutator quadratic in the 
generators and antisymmetric:
\eqa
[P_\mu,P_\nu]_{_\FF}&=&[P_\mu,P_\nu]\,,\nn\\[.3em]
[P_\rho , M^\FF_{\mu\nu}]_{_\FF}&=&[P_\rho , M^\FF_{\mu\nu}]\,,\nn\\[.3em]
[M^\FF_{\mu\nu},M^\FF_{\rho\sigma}]_{_\FF}&=&
[M^\FF_{\mu\nu},M^\FF_{\rho\sigma}]
-i\theta^{\al\be}[P_\al,M_{\rho\sigma}][P_\be,M_{\mu\nu}]~.
\ena
From (\ref{LiePoincF}) we immediately obtain the Jacoby identities:
\eq
[t \,,[t',t'']_{_\FF} ]_{_\FF} +[t' \,,[t'',t]_{_\FF} ]_{_\FF} 
+ [t'' \,,[t,t']_{_\FF} ]_{_\FF} =0~,
\en
for all $t,t',t''\in iso^{\FF\!}(3,1)$.
\sk

\sk\noi
{\bf Acknowledgements}
I would like to thank the organizers and the participants of the second Modave Summer 
School in Mathematical Physics for the
fruitful and refreshing week, characterized by
lectures, discussions, group study, all in a community atmosphere.

This work is partially supported by the 
European Community Marie Curie contracts MERG-CT-2004-006374 (European Reintegrtion Grant) and MRTN-CT-2004-005104 (Human Potential programme), and 
by the Italian MIUR contract PRIN-2005023102.

\appendix
\section{Algebras, Coalgebras and Hopf algebras}
In the introduction we have motivated the notion of Hopf algebra.
We here review some basic definitions in linear algebra and show how 
Hopf algebras merge algebras and coalgebras structures in a symmetric 
(specular) way ($\!\!$\cite{Sweedler}). 
\sk
We recall that a module by definition is an abelian group. 
The group operation is denoted $+$ (additive notation).
A vector space $A$ over $\cc$ (or $\rr$) is a $\cc$-module, 
i.e. there is an action $(\la,a)\rightarrow 
\la a$ of the group $(\cc-\{0\},\cdot)$ on the module $A$,
\eq
(\la'\la) a = \la(\la' a)~,
\en 
and this action is compatible with 
the addition in $A$ and in $\cc$, i.e. it is compatible with the module 
structure of $A$ and of $\cc$:
\eq
\la(a+a')=\la a+\la a'~~,~~~(\la+\la')a=\la a+\la' a~.
\en
An {\bf algebra} $A$ over $\cc$ with unit $I$, is a vector space over $\cc$
with a multiplication map, that we denote $\cdot$ or $\mu$, 
\eq
\mdot \,:\, A\times A\rightarrow A
\en
that is $\cc$-bilinear: $(\la a)\cdot (\la' b)=\la\la'(a\cdot b)$,
that is associative and that for all $a$ satisfies $a\cdot I=I\cdot a=a$.

These three properties can be stated diagrammatically.
$\cc$-bilinearity of the product $\mdot : A\times A\rightarrow A$,
is equivalently expressed as linearity of the map 
$\mdot : A\otimes A\rightarrow A$. Associativity reads,
\[
\begin{CD}
{A\otimes A\otimes A}                   @>\mdot\otimes id>>              {A\otimes A}\\
 @V id\otimes \mdot VV                                                               @V\mdot VV\\
{A\otimes A}@>\mdot >>{A}
\end{CD}
\]
Finally the existence of the unit $I$ such that for all $a$ we have
$a\cdot I=I\cdot a=a$ is equivalent to the existence of a linear map
\eq
i : \cc\rightarrow A
\en
such that
$$
\begin{CD}
{\cc\otimes A}                   @> i\otimes id>>              {A\otimes A}\\
 @V \simeq VV                                                               @V\mdot VV\\
{A}@>id >>{A}
\end{CD}
$$
and

$$
\begin{CD}
{A\otimes \cc}                   @> id\otimes i>>              {A\otimes A}\\
 @V \simeq VV                                                               @V\mdot VV\\
{A}@>id >>{A}
\end{CD}
$$

\noi where $\simeq$ denotes the canonical isomorphism between $A\otimes \cc$ 
(or $\cc\otimes A$) and $A$.
The unit $I$ is then recovered as $i(1)=I$.
\sk

A {\bf coalgebra}  $A$ over $\cc$ is a vector space with a linear 
map $\D : A\rightarrow A\otimes A$ that is coassociative, 
$(id \otimes \D)\D=(\D\otimes id)\D$, and a linear map 
$\epsi : A\rightarrow \cc$, called counit that satisfies
$(id\otimes \epsi)\D(a)=(\epsi\otimes id)\D(a)=a$. These properties can be 
expressed diagrammatically by reverting the arrows of the previous diagrams:
$$
\begin{CD}
{A\otimes A\otimes A}                   @<\D\otimes id<<              {A\otimes A}\\
 @A id\otimes \D AA                                                               @A\D AA\\
{A\otimes A}@<\D <<{A}
\end{CD}
$$

$$
\begin{CD}
{\cc\otimes A}                   @< \epsi\otimes id<<              {A\otimes A}\\
 @A \simeq AA                                                               @A\D AA\\
{A}@<id <<{A}
\end{CD}
$$
and

$$
\begin{CD}
{\cc\otimes A}                   @< id\otimes \epsi<<             {A\otimes A}\\
 @A \simeq AA                                                               @A\D AA\\
{A}@<id <<{A}
\end{CD}
$$

\sk
We finally arrive at the 
\sk
\noi {\sl Definiton} A {\bf bialgebra} $A$ over $\cc$ is a vectorspace $A$ with 
an algebra and a coalgebra structure that are compatible, i.e. 

\noi 1) the coproduct 
$\D$ is an algebra map between the algebra $A$ and the algebra $A\otimes A$,
where the product in $A\otimes A$ is 
$(a\otimes b)(c\otimes d)=ac\otimes bd$,
\eq\label{appcop}
\D(ab)=\D(a)\D(b)~~~,~~\D(I)=I\otimes I
\en
2) The counit $\epsi : A\rightarrow \cc$ is an algebra map 
\eq\label{appepsi}
\epsi(ab)=\epsi(a)\epsi(b)~~~,~~\epsi(I)=1~.
\en
\sk
\noi {\sl Definition} A {\bf Hopf algebra} is a bialgebra with a linear map $S : A\rightarrow A$, called antipode (or coinverse), such that
\eq\label{selfant}
\mdot(\kappa\otimes id)\D(a)=\mdot(id\otimes\kappa)\D(a)=\epsi(a) I ~.
\en
\noi
It can be proven that the antipode $S$ is unique and antimultiplicative  
$$S(ab)=S(b)S(a)~.$$
\sk
From the definition of bialgebra it follows that 
$\mu : A\otimes A\rightarrow A$ 
and $i : \cc\rightarrow A$ are coalgebra maps, i.e., 
$\D\circ\mu=\mu\otimes \mu\circ\underline{\D}$, 
$~\epsi\otimes \mu=\underline\epsi$ 
and $\D\circ i=i\otimes i\circ \D_\cc$, $~\epsi\circ i =\epsi_\cc\,,$ 
where the coproduct and counit in $A\otimes A$ are given by 
$\underline{\D}(a\otimes b)=a_1\otimes b_1\otimes a_2\otimes b_2$ and 
$\underline{\epsi}=\epsi\otimes\epsi$, while the
coproduct in $\cc$ is the map $\D_\cc$ that identifies $\cc$ with 
$\cc\otimes\cc$ and the counit is $\epsi_\cc=id$.  Viceversa 
if $A$ is an algebra and a coalgebra and $\mu$ and $i$ are coalgebra maps then it follows that $\D$ and $\epsi$ are algebra maps.
\sk
One can write diagrammatically equations 
(\ref{appcop}), (\ref{appepsi}),  (\ref{selfant}), and see that the Hopf algebra 
definition is invariant under inversion of arrows and exchange 
of structures with costructures, 
with the antipode going into itself. In this respect the algebra and the 
coalgebra structures in a Hopf algebra are specular.

\end{document}